# RECENT RESULTS OF NON-ACCELERATOR-BASED NEUTRINO EXPERIMENTS

YIFANG WANG[*]

*Institute of High Energy Physics*
*Beijing, 100049, P.R. China*

Recent results of non-accelerator-based experiments, including those of solar, atmospheric, and reactor neutrinos oscillations, neutrinoless double-beta decays, and neutrino magnetic moments, are reviewed. Future projects and their respective prospects are summarized.

## 1. Introduction

Neutrino physics is a focus of particle physics, astrophysics and cosmology due to its being the end products of decays of almost all particles, its abundance in the universe and its peculiar intrinsic properties. Its mass is crucial for the universe, its evolution and the large structure formation.

Although neutrinos are discovered for more than half a century, we still know very little about their intrinsic properties, such as the mass and the magnetic moments, due to their extremely weak interactions with matter. Since the observation of neutrino oscillations by the Super-Kamiokande[1] experiment using atmospheric neutrinos, significant progress have been made and a flood of new results from SNO[3] and KamLAND[4], in addition to Super-Kamiokande[2], were reported.

In this talk, I will review recent results on the study of neutrino oscillations, the absolute neutrino masses, and the neutrino magnetic moments from non-accelerator-based experiments in the last two years. For accelerator-based experiments, please refer to McGrew's talk[5] of this conference.

## 2. Neutrino Oscillations

First proposed by Pontecorvo, oscillation is a quantum phenomena if the mass and the weak eigenstates of neutrinos are not identical. The transformation of different eigenstates for the three-generation of active neutrinos can be described as the following:

$$\begin{pmatrix} \nu_e \\ \nu_\mu \\ \nu_\tau \end{pmatrix} = \begin{pmatrix} V_{e1} & V_{e2} & V_{e3} \\ V_{\mu 1} & V_{\mu 2} & V_{\mu 3} \\ V_{\tau 1} & V_{\tau 2} & V_{\tau 3} \end{pmatrix} \begin{pmatrix} \nu_1 \\ \nu_2 \\ \nu_3 \end{pmatrix}$$

This matrix, often called Pontecorvo-Maki-Nakagawa-Sakata(PMNS) matrix[6], can be also written as,

$$V = \begin{pmatrix} 1 & 0 & 0 \\ 0 & c_{23} & s_{23} \\ 0 & -s_{23} & c_{23} \end{pmatrix} \begin{pmatrix} c_{13} & 0 & s_{13} \\ 0 & e^{-i\delta} & 0 \\ -s_{13} & 0 & c_{13} \end{pmatrix} \begin{pmatrix} c_{12} & s_{12} & 0 \\ -s_{12} & c_{12} & 0 \\ 0 & 0 & 1 \end{pmatrix} \begin{pmatrix} e^{i\rho} & 0 & 0 \\ 0 & e^{i\sigma} & 0 \\ 0 & 0 & 1 \end{pmatrix}$$

Such a representation is a product of four matrices: the first one responsible for the atmospheric neutrino oscillation, the third one for the solar neutrino oscillation and the second one for the mixture of the two. The last one containing Majorana phases only appears in neutrinoless-double-beta decays. The three-generation neutrino transformation is hence completely described by two mass differences: $\Delta m^2_{23}$, $\Delta m^2_{12}$, three mixing angles: $\theta_{12}$, $\theta_{23}$, $\theta_{13}$, and one CP phase $\delta$, plus two Majorana phases.

Neutrino oscillation is important since it provides an experimental tool to access extremely small neutrino masses. For two-flavor oscillation in vacuum, the oscillation probability is expressed as:

$$P(\nu_1 \to \nu_2) = \sin^2 2\theta \sin^2(1.27\Delta m^2 L/E).$$

Here $\sin^2 2\theta$ denotes the oscillation amplitude and $(1.27\Delta m^2 L/E)$ represents the oscillation frequency. Currently there are three evidences of neutrino oscillations, demonstrated in the oscillation parameter space of $\sin^2 2\theta$ and $\Delta m^2$, as shown in Fig.1. While both solar and atmospheric neutrino oscillations have been observed by several experiments, the so-called LSND effect has not been confirmed yet[7]. For details, please refer to Ref.[8,9] for reviews of neutrino oscillations and Ref.[10] for a global fit of oscillation parameters.

---

[*] yfwang@ihep.ac.cn

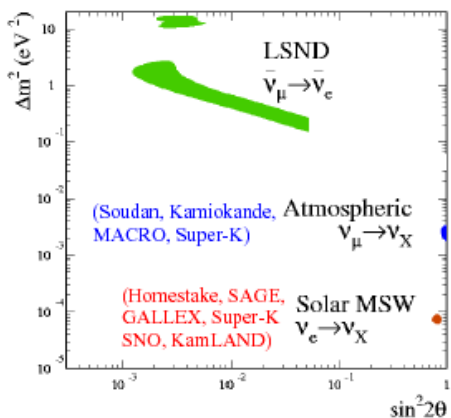

Figure 1 Evidence of neutrino oscillation: allowed regions in the oscillation parameter space from all experiments.

### 2.1. *Atmospheric neutrinos*

Atmospheric neutrinos originate from cosmic-ray-induced showers in the atmosphere at high altitude, producing typically two $\nu_\mu$ and one $\nu_e$ from pion decays. The so-called "atmospheric neutrino anomaly", a deficit of neutrinos with respect to the naïve expectation ($\nu_\mu/\nu_e=2$), was observed by numerous experiments, sometimes also controversial[8]. The Super-Kamiokande experiment showed unambiguously that the deficit of $\nu_\mu$ is the cause of the anomaly and the data are consistent with neutrino oscillation of $\nu_\mu$ to $\nu_\tau$[1,2]. This result is confirmed by others although with less statistical significance[8].

The Super-Kamiokande detector consists of 50,000 t of pure water and 10,000 20" PMTs. Neutrinos are observed via the Cerenkov rings of charged leptons produced from either the charged current interactions or ν-e elastic scattering. New analysis results of full Super-K data before the accident in 2001(SK-I) were reported in this conference[11]. Several improvements are incorporated including a new 3D calculation of neutrino flux, new neutrino interaction parameters tuned by K2K data and a new $\chi^2$ calculation to treat each systematic error independently. The zenith angle distributions of e-like and μ-like events are shown in Fig.2.

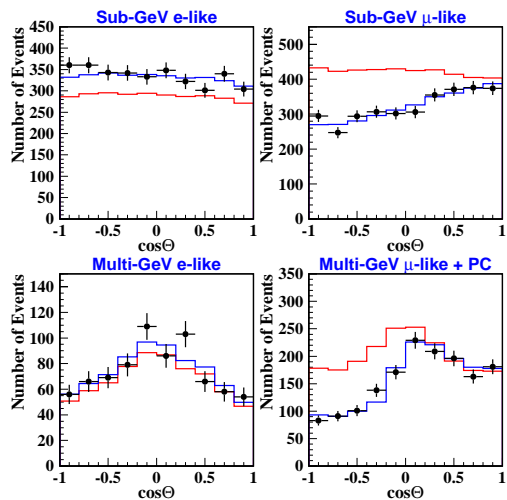

Figure 2 Zenith angle distribution of e-like and μ-like events from the Super-K experiment[11]. The red line is for null oscillation and the blue line is the best fit assuming $\nu_\mu$ to $\nu_\tau$ oscillation

A clear deficit of $\nu_\mu$ events, particularly of those traveling through the earth with longer baseline ($\cos\theta \sim -1$), can be seen from the figure. A 2-flavor analysis of $\nu_\mu$ to $\nu_\tau$ oscillation, taking into account the matter effect, results in an allowed region for oscillation parameters as shown in Figure 3. The best fit gives $\sin^2 2\theta = 1.0$ and $\Delta m^2 = 2.1 \times 10^{-3}$ eV$^2$, while at 90% C.L., $\sin^2 2\theta > 0.92$ and $1.5 < \Delta m^2 < 3.4 \times 10^{-3}$ eV$^2$.

The oscillation parameters are further constrained by selecting events with good resolution of L/E since it determines the oscillation frequency. Figure 4 shows the ratio of data to the prediction without oscillation as a function of L/E, compared to the best fit of 2-flavor oscillation. Two alternative explanations, neutrino decay[12] and neutrino decoherence[13] are also shown. It can be seen clearly from the plot that neutrino oscillation explains better the data, particularly the characteristic dip at L/E ~ $5\times10^2$. Neutrino decay and neutrino decoherence are disfavored at $3.4\sigma$ and $3.8\sigma$ respectively.

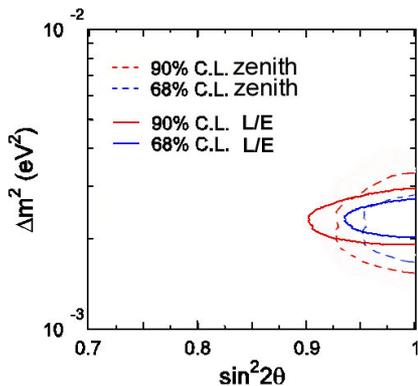

Figure 3 Allowed region for 2-flavor oscillations of $\nu_\mu$ to $\nu_\tau$ from both the zenith angle and L/E analysis.

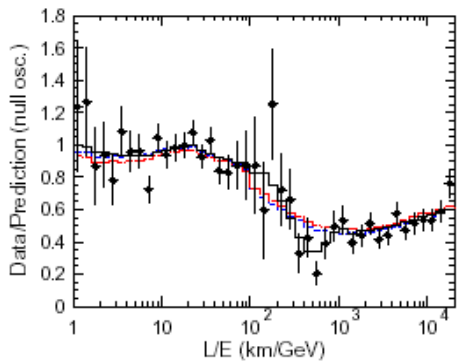

Figure 4 Ratio of data to the prediction without oscillation as a function of L/E, compared to the best fit of 2-flavor oscillations(solid line), neutrino decay (dashed line) and neutrino decoherence (dotted line).

A 2-flavor oscillation analysis leads to a smaller allowed region for $\Delta m^2$, reflecting its better sensitivity due to its better L/E resolution, as shown in Fig.3. The best fit yields $\sin^2 2\theta$ = 1.0, $\Delta m^2 = 2.4 \times 10^{-3}$ eV$^2$, while at 90% C.L., $\sin^2 2\theta > 0.90$ and $1.9 < \Delta m^2 < 3.0 \times 10^{-3}$ eV$^2$.

A three-flavor analysis was also performed by the Super-Kamiokande experiment, which leads a best fit of $\sin^2 2\theta_{23} = 1.0$, $\sin^2 2\theta_{13} = 0.0$, and $\Delta m^2 = 2.7 \times 10^{-3}$ eV$^2$.

New data after the accident from the Super-K experiment(SK-II) are collected with good quality, and consistent with that before the accident. We expect to see results in a not too far future.

## 2.2. *Solar neutrinos*

Solar neutrinos are produced from fusion reactions in the sun together with energies in the form of light and heat shining the sky. The Standard Solar Model(SSM) predicted the electron neutrino flux and the energy spectrum with a great precision, no other types of neutrinos can be produced directly from the sun[14].

Since 70's the observed solar neutrinos from all the experiments are only 1/3-2/3 of what the Standard Solar Model predicted, known as the "Solar neutrino Puzzle"[8]. The break through in 2001 by the SNO experiment occurred after the observation of neutrinos other than the electron type, a clear evidence of flavor conversion, due most probably to neutrino oscillation[3]. The idea of using the heavy water, $D_2O$, to detect all three types of neutrinos was proposed in 1985 by Herb Chen[15], who unfortunately passed away in 1987, long before the great success of SNO. In his proposal, the electron neutrino can be detected via the charged current(CC) reaction, $\nu_e + d \rightarrow e^- + p + p$, through the Cerenkov ring from electrons. The neutral current(NC) reaction, $\nu_x + d \rightarrow \nu_x + n + p$, reveals the neutrino flux of $\nu_x = \nu_e + \nu_\mu + \nu_\tau$ through the detection of neutrons. The event rate of elastic scattering(ES), $\nu_x + e^- \rightarrow \nu_x + e^-$, actually proportional to $\nu_x = \nu_e + (\nu_\mu + \nu_\tau)/6$, are obtained via the detection of electrons. Neutrino fluxes of $\nu_e$, $\nu_\mu$, $\nu_\tau$ can therefore be measured by the three reactions unambiguously.

The Phase II SNO results using NaCl to enhance the neutron detection, instead of the neutron capture on Deuterium in Phase I, are reported[16]. The distributions of the event isotropy parameter $\beta_{14}$, the event direction relative to the vector from the sun, and the kinetic energy are shown Fig. 5. The number of CC, NC and ES events, obtained from a fit to these plots plus the distribution of events as a function of radius, is listed in the following:

$$\Phi_{CC} = 1.76^{+0.06}_{-0.05}(\text{stat})^{+0.09}_{-0.09}(\text{sys})$$
$$\Phi_{ES} = 2.39^{+0.24}_{-0.23}(\text{stat})^{+0.12}_{-0.12}(\text{sys})$$
$$\Phi_{NC} = 5.09^{+0.44}_{-0.43}(\text{stat})^{+0.46}_{-0.43}(\text{sys})$$

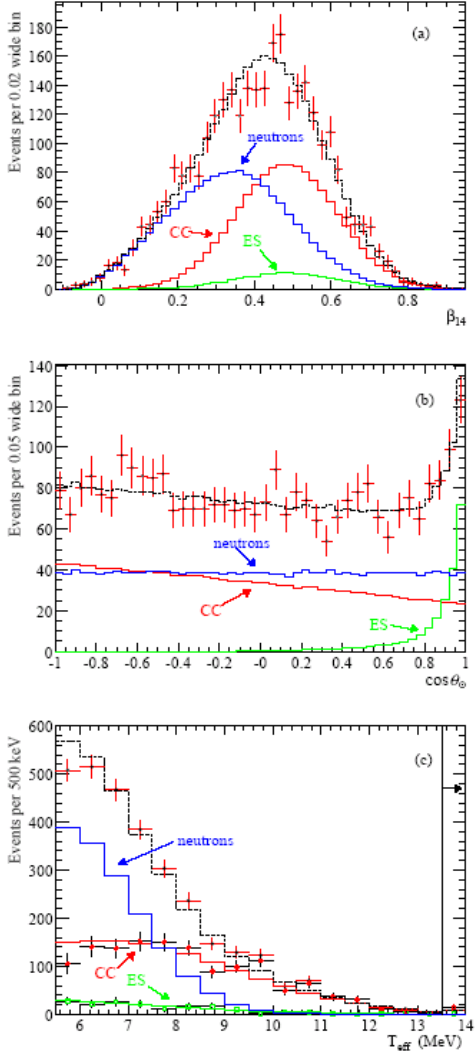

Figure 5 Distribution of (a) isotropy, (b) polar angle relative to the vector from the sun, (c) kinetic energy for the selected events. The CC, NC, ES and background events are extracted from the fit to these plots together with event distribution as a function of radius. The dashed line represents the sum of all components.

These numbers are consistent with Phase I results at a similar precision[3], and neutrino fluxes of $\nu_e$ and $\nu_\mu + \nu_\tau$ can be derived as shown in Fig. 6. Although exotic solutions such as spin-flip and other non-standard interactions[17] can not be excluded completely, neutrino oscillation is the most favorable solution to the fact of $\phi_{\mu\tau} > 0$.

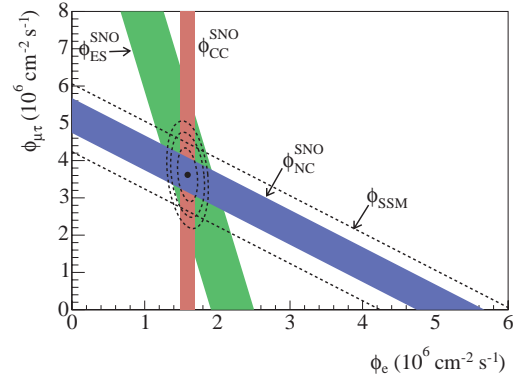

Figure 6 Neutrino fluxes measured by the SNO experiment using the NaCl salt phase data.

A combined fit[16], utilizing the current available data of solar neutrinos from Super-K, SNO, Homestake, Gallex and Sage, together with the SSM prediction on the $\nu_e$ flux and energy spectrum without fixing the normalization, finds that data are consistent with the SSM prediction assuming neutrino oscillation. The allowed region of the mixing parameters is shown in Fig. 7, and the best fit yields $\tan^2\theta = 0.398$, $\Delta m^2 = 6.46\times10^{-5}$ eV$^2$.

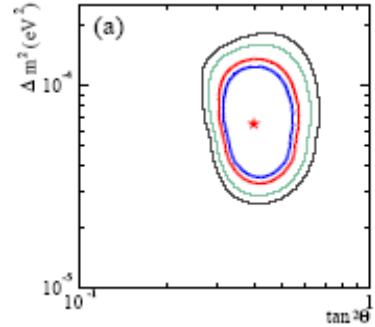

Figure 7 The allowed neutrino mixing parameter space from SNO and other solar neutrino experiments.

SNO is now on their phase III to detect neutrons using $^3$He tube. A total of 40 proportional counters will be deployed this year and new data is expected to come soon.

The Super-Kamiokande experiment reported a new analysis of full SK-I solar neutrino data[18]. By using an un-binned analysis, taking into account energy and zenith angle dependence of event rate variation, the new day/night effect is measured to be $A_{ND} = -0.018 \pm 0.016^{+0.013}_{-0.012}$, in consistent with their previous results[19].

## 2.3. Reactor neutrinos

Reactor neutrinos are produced through fission of nuclear fuels containing mainly four isotopes, $^{238}$U, $^{235}$U, $^{239}$Pu and $^{241}$Pu, and the subsequent decays. Typically, one fission reaction will produce 200 MeV energy and 6 electron anti-neutrinos. Hence a reactor with 3 GW thermal power produces $6\times10^{20}$ neutrinos per second, an extremely powerful, clean neutrino source.

Reactor-based neutrino oscillation experiments are so-called "disappearance experiment", which looks for the deficit of neutrino flux and the distortion of neutrino energy spectrum with respect to the expectation. The flux depends strictly on the thermal power of the reactor, which can be measured up to 0.7% precision. The energy spectrum can be predicted by three methods: (1) theoretical calculation of beta energy spectra of all isotopes and their decay products in the core; (2) measurement of beta spectra of three isotopes, $^{238}$U, $^{239}$Pu and $^{241}$Pu, and their decay products, combined with a theoretical calculation for $^{235}$U; (3) previous measurement of the neutrino energy spectrum at very short baselines. Because the method (2) and (3) are consistent within 2.0%, it is generally believed that the prediction of the reactor neutrino flux and the spectrum combined is within 3% precision[20].

Since 80's reactor-based experiments have consistently observed neutrino signals in agreement with that of the expectation until the KamLAND experiment which for the first time, found the neutrino deficit using a man-made source[4]. KamLAND is a 1kt liquid scintillator detector shielded by water and mineral oil.

Electron anti-neutrinos are detected via the inverse beta decay $\nu_e + p \rightarrow e^+ + n$ with an energy threshold of 1.8 MeV. The cross section of this process is well known via the neutron lifetime[21]. The cross-section-weighted neutrino energy, peaked at ~ 4 MeV, can be derived from that of the measured prompt positron signal. Neutrons are captured on protons in the liquid scintillator, releasing a 2.2 MeV γ-ray as a delayed signal with a time constant of 180 μs. Such a two-fold coincidence is crucial to suppress backgrounds from cosmic-rays and environmental radioactivity.

KamLAND reported a much improved results with higher statistics[22,23]. Fig. 8 shows the measured prompt energy spectrum in comparison with the expectation for no oscillations and the best fit assuming oscillation. The measured spectrum is clearly distorted with respect to that of no oscillation and a simple re-normalization of the neutrino flux can not explain it.

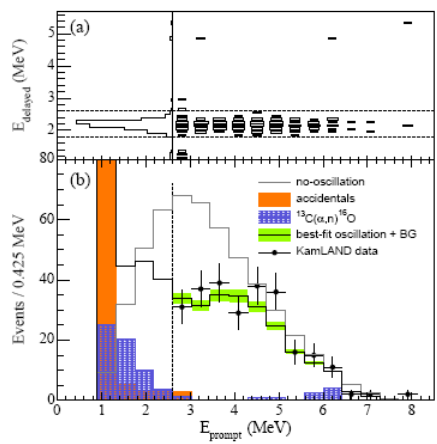

Figure 8 a) The correlation of energies between the prompt and delayed events after cuts. The three events with $E_{delayed}$ ~ 5 MeV are consistent with neutron capture on Carbon. b) Prompt energy spectrum of neutrino candidates and background spectra. The shaded band indicates the systematic error in the best-fit reactor spectrum above 2.6 MeV.

As discussed in session **2.1**, oscillation is generally best demonstrated in terms of L/E if statistics is sufficient and resolution is good enough. Fig. 9 shows the ratio of the observed data to the expectation without oscillation, as a function of $L_0/E$, where $L_0$=180km. The best-fit assuming 2-flavor oscillation, together with the prediction of neutrino decay[12] and decoherence[13] are also shown. Clearly oscillation is favored, similar to the case of atmospheric neutrinos.

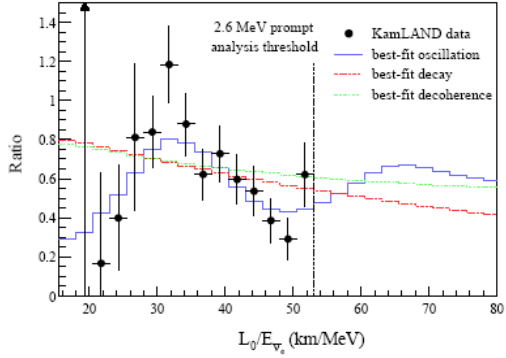

Figure 9 Ratio of the observed anti-neutrino spectrum to the expectation for no-oscillation as a function of $L_0/E$, where $L_0$=180 km. The best-fit assuming 2-flavor oscillation, together the prediction neutrino decay and decoherence are also shown.

A 2-flavor analysis of neutrino oscillation using all the KamLAND data, taking into account detailed reactor information, results in allowed regions(shaded region) of mixing parameters as shown in Fig. 10a[23]. Results from all solar neutrino experiments(lines), taking into account the SSM prediction[16], are also shown. A combined 2-flavor oscillation analysis of solar neutrinos and reactor anti-neutrinos, assuming CPT invariance, yields an allowed region in the oscillation parameter space as shown in Fig. 10b. The sensitivity to $\Delta m^2_{12}$ is dominated by the observed distortion of the KamLAND spectrum, while the solar neutrino data provide the best constraint to $\theta_{12}$. The best fit of the combined analysis is $\Delta m^2_{12} = 7.9^{+0.6}_{-0.5} \times 10^{-5} eV^2$ and $\tan^2\theta = 0.40^{+0.10}_{-0.07}$.

The spin-flip and other non-standard interactions as the explanation of the flavor conversion of solar neutrinos[17] is ruled out by the KamLAND observation.

## 2.4. *Future prospects of neutrino oscillations*

Since experimental results from Super-K, SNO and KamLAND all disfavor alternative explanations, neutrino oscillations are established. Among 6 oscillation parameters, we know now $\Delta m^2_{21}$, $\sin^2 2\theta_{12}$, $|\Delta m^2_{32}|$ and $\sin^2 2\theta_{23}$, and we are in quest of the rest: $\sin^2 2\theta_{13}$, CP phase δ and the sign of $\Delta m^2_{32}$. The CP phase term only appears in the form of $\sin\delta \cdot \sin\theta_{13}$, hence the measurement of $\sin\theta_{13}$, based either on reactor or long baseline accelerator experiments has the highest priority.

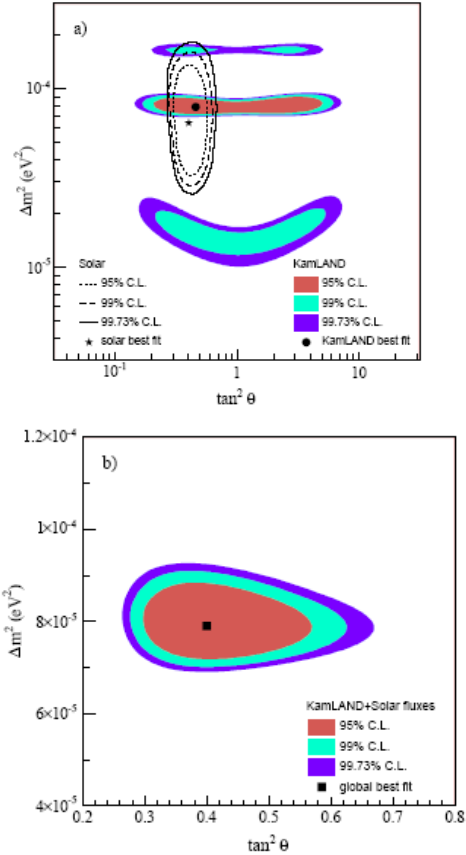

Figure 10 a) Allowed regions of neutrino oscillation parameters from KamLAND anti-neutrino data (shaded regions) and solar neutrino experiments (lines); b) Results of a combined two-flavor fit of KamLAND and solar neutrino results assuming CPT invariance.

The advantages to measure $\sin^2 2\theta_{13}$ at reactors are obvious: the signal is clean without cross-talk with matter effects and CP phase; the experiment is relatively cheap and can be quickly deployed in principle. Hence it provides the direction for future accelerator experiments.

The amplitude of $\sin^2 2\theta_{13}$ is constrained by the results from Chooz[24], namely $\sin^2 2\theta_{13} <$ 0.15 at 90% C.L. for $\Delta m^2_{23} = 2.0 \times 10^{-3}$ eV$^2$.

Global fits of all experimental results show that $\sin^2 2\theta_{13} \sim 0.03$[25], or $\sin^2 2\theta_{13} < 0.09$ at 90% C.L.[10]. It is generally desired to reach the precision at 1% level for the future $\sin^2 2\theta_{13}$ measurement[26].

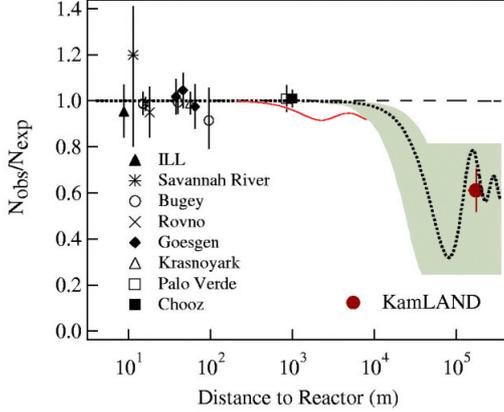

Figure 11 Ratio between the observed number of neutrino events and that of the expected one assuming no oscillation as a function of baseline for all the past reactor neutrino experiments. The best fit assuming oscillation(solid line), together with the sub-dominant effect of $\theta_{13}$, assuming $\sin^2 2\theta_{13}=0.1$(red line), are also shown.

Previous reactor experiments[4,20], as shown in Fig. 11, have a typical precision of (3-6)%, in which, (2-3)% is reactor related, (1-3)% backgrounds related, and (1-4)% detector related. These errors may be reduced to the desired level by following ways: 1) use two detectors, near and far, to cancel reactor related errors[27]; 2) use movable detectors between the near and far site to cancel most of the detector related errors; 3) construct the detector with enough shielding and deep underground to eliminate backgrounds from environmental radioactivity, cosmic-ray-induced $^8$He/$^9$Li and neutrons[28]; 4) find an optimum baseline; 5) establish a comprehensive calibration program and pay a close attention to all the details of the experiment, such as the energy threshold, resolution, scintillator purity and transparency, etc. Of course, the integrated luminosity(reactor power × target mass × running time)play a vital role on the statistical error.

Currently there are several proposals[26], as listed in Table 1 and details can be found in Ref.[29]. Fig. 12 shows the comparison of sensitivities of some of the proposals, in a way similar to that used in Ref.[30] as a function of integrated luminosity.

Among all the proposals, the Double Chooz experiment in France has been approved and they expect to take data in 2008. All others are in different stages of funding process. The Daya Bay experiment in Shenzhen, China has the highest luminosity and the expected sensitivity is 1% at 90% C.L. after 3 years running.

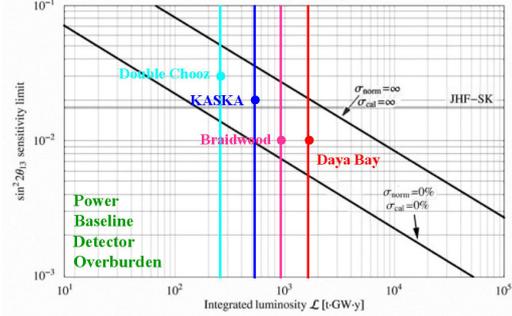

Figure 12 Expected sensitivity(from their own claim) of proposed $\theta_{13}$ experiments as a function of integrated luminosity. Some of the important factors, such as baseline and overburden, are not demonstrated. The luminosity did not normalize to the same baseline.

Table 1. Comparison of parameters of all proposed $\theta_{13}$ experiments.

| Site (Proposal) | Power (GW) | Baseline Near/Far (m) | Detector mass Near/Far (t) | Overburden (MWE) | Sensitivity (90%C.L.) |
|---|---|---|---|---|---|
| Angra Dos Reis | 4.1 | 300/1300 | 50/500 | 200/1700 | 0.007 |
| Braidwood (USA) | 6.5 | 270/1800 | 25/50 | 450/450 | 0.01 |
| Double Chooz (France) | 8.4 | 150/1050 | 10/10 | 60/300 | 0.03 |
| Daya Bay (China) | 11.6 | 350/1800 | 20/40 | 250/1200 | 0.01 |
| Diablo Canyon(USA) | 6.4 | 400/1800 | 25/50 | 100/700 | 0.01 |
| Kashiwazaki(Japan) | 24.3 | 350/1300 | 8.5/8.5 | 300/300 | 0.02 |
| Krasnoyarsk(Russia) | 3.2 | 115/1000 | 46/46 | 600/600 | 0.01 |

## 3. Absolute neutrino masses

Up to now we discussed neutrino oscillations with which only neutrino mass differences can be measured. Absolute neutrino masses are usually searched for via Tritium decays, with the endpoint $E_0$=18.574 KeV. The neutrino mass in beta decays is actually related to weak eigenstates through the PMNS matrix, $m_{\nu e}=[\Sigma_i|U_{ei}|^2 m^2_{\nu i}]^{1/2}$. Currently the best limit is $m_{\nu e}$<2.2 eV at 95% C.L. from Mainz and Troitsk experiments[31]. The Katrin experiment, expected to be online in 2006, can reach a sensitivity of about 0.2 eV at 95% C.L., a limit interested to astrophysics and cosmology[32].

$^3H \rightarrow {}^3He + e^- + \nu_e$,

Recent results from WMAP and other experiments show that neutrino masses are strongly limited, or even finite, as listed in table 2. Different limits in the table result from different assumptions, fitting procedures and datasets. Clearly these limits are more stringent than those from Tritium decays and are already a strong constraint to the unconfirmed LSND[7] and Heidelburg-Moscow $\beta\beta$ decay[33] results.

Table 2. Current results on the neutrino mass limit from astrophysics experiments.

| Data | $\Sigma m_i$ (eV) @95%CL | Ref. |
|---|---|---|
| 2dFGRS | <1.8 | [34] |
| WMAP+2dF+… | <0.7 | [35] |
| WMAP+2dF | <1.0 | [36] |
| XLF+WMAP+2dF+… | $0.56^{+0.3}_{-0.26}$ | [37] |
| SDSS+WMAP | <1.7 | [38] |
| WMAP+ACBAR+ 2dF+SDSS | <1.6 | [39] |

## 4. Neutrinoless $\beta\beta$ decays

The neutrinoless double beta decay[40],

$(A, Z) \rightarrow (A, Z+2) + 2e^-$,

is a hypothetical process if neutrinos are massive and are of Majorana nature, namely, identical to their anti-particles. The neutrino mass to be observed can be expressed as $<m_{ee}>=[\Sigma_i(U_{ei})^2 m^2_{\nu i}]^{1/2}$, similar to that of beta decays. The $\beta\beta(0\nu)$ decay is distinctive by its monochromatic $\beta$ spectrum, while that of the conventional $\beta\beta(2\nu)$ decay, a second order nuclear process, is continuous. Thus the detector resolution and environmental backgrounds are extremely important for such a rare signal experiment.

Numerous searches for the $\beta\beta(0\nu)$ decays have been performed[8], only one claimed a positive signal[33]. In this conference, first results from NEMO-3[41] are reported, as listed in Table 3 together with results from other experiments.

Table 3. Some of the current results and future prospects of double beta decay experiments

| Current experimental limits | | | |
|---|---|---|---|
| | Isotopes | Kg*y | $<m_\nu>$(eV) (90%CL) |
| H-M[33] | $^{76}$Ge | 30 | 0.24–0.58 |
| NEMO-3[41] | $^{82}$Se | 0.5 | <1.3–3.6 |
| NEMO-3[41] | $^{100}$Mo | 4.1 | <0.7–1.2 |
| Xe TPC[42] | $^{136}$Xe | 10 | <2.4–2.7 |
| [43] | $^{130}$Te | 3 | <1.1–2.6 |
| Sensitivity of future experiments | | | |
| | Isotopes | Mass (t) | $<m_\nu>$ (eV) (90%CL) |
| CUORE[44] | $^{130}$Te | 0.75 | ~0.03 |
| EXO[45] | $^{136}$Xe | 10.0 | ~0.01 |
| GENIUS[46] | $^{76}$Ge | 1.0 | ~0.01 |
| Majorana[47] | $^{76}$Ge | 0.42 | ~0.02 |
| MOON[48] | $^{100}$Mo | 3.0 | ~0.01 |
| Super-NEMO[41] | $^{82}$Se | >0.1 | ~0.03 |

R&D results of the EXO[45] and XMASS[49] experiments for $\beta\beta$ decays are reported in this conference. EXO observed $^{136}Ba^+$ in a 0.01 Torr Xenon gas for an indefinite lifetime using a novel laser tagging technology. This is their first step towards the identification of $^{136}Ba^{++}$ from $\beta\beta$ decays in a liquid or a highly pressured gas, in order to battle non-$\beta\beta(2\nu)$ backgrounds. An incomplete list of future $\beta\beta$ decay experiments is listed in Table 3. Some of them will be realized and hopefully, the effective neutrino mass will be pushed to the level of ~0.01 eV.

## 5. Neutrino magnetic moments

In the Standard Model, a finite neutrino mass will lead to a finite neutrino magnetic moment at the level of $\mu_\nu \sim 3.2\times10^{-19}\,(m_\nu/1\text{ eV})$. It can be enhanced significantly by new physics beyond the Standard Model. Texono experiment reported their latest results[50] by searching for excess in $\nu$–e scattering whose spectrum follows $(1/T-1/E_\nu)\,\mu_\nu^2$. No excess down to 12 KeV during the reactor on and off periods are observed and the up limit of the magnetic moment at 90% C.L. is set to be $1.3\times10^{-10}\,\mu_B$, as shown in Fig. 13, together with results from other experiments.

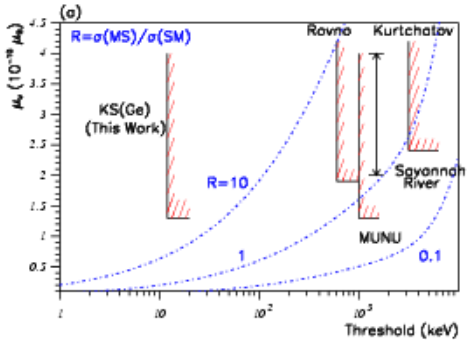

Figure 13 Bounds of the neutrino magnetic moment as a function of their experimental threshold.

The Super-Kamiokande and Borexino experiments have searched for solar neutrino magnetic moments by looking at $\nu$-e scattering spectrum and observed no excess. Similar results are obtained also from LSND and DONUT experiment as listed in table 4. Future experiments can reach the level of $\sim10^{-12}\,\mu_B$, still a bit far away from the Standard Model prediction, but the nature may give us a surprise.

## 6. Summary

Neutrino oscillations have been established based on results from solar, atmospheric and reactor experiments. The solar mixing angle is large but not maximal while the atmospheric mixing angle is maximal. The two mass differences of three-generation neutrinos are fairly known.

Our next goal is to know the absolute neutrino mass, mostly from Tritium decays, the Majorana or Dirac nature of neutrinos from neutrinoless double beta decays, the mixing angle $\theta_{13}$ from reactors and ultimately, the CP phase and the mass hierarchy from very long baseline neutrino oscillation experiments, presumably based on accelerators.

Table 4 Current results and future prospects of neutrino magnetic moments

| Current Experimental limits | | |
|---|---|---|
| Experiments | Up limits (90%CL) | Ref. |
| Texono | $\mu_\nu(\nu_e) <1.3\times10^{-10}\,\mu_B$ | [50] |
| MUNU | $\mu_\nu(\nu_e) <1.0\times10^{-10}\,\mu_B$ | [51] |
| SK+ $\nu_\odot$+ KamLAND | $\mu_\nu(\nu_\odot) <1.1\times10^{-10}\,\mu_B$ | [52] |
| Borexino | $\mu_\nu(\nu_\odot) <5.5\times10^{-10}\,\mu_B$ | [53] |
| LSND | $\mu_\nu(\nu_e) <1.1\times10^{-9}\,\mu_B$ $\mu_\nu(\nu_\mu) <6.8\times10^{-10}\,\mu_B$ | [54] |
| DONUT | $\mu_\nu(\nu_\tau) <3.9\times10^{-7}\,\mu_B$ | [55] |
| Future Experiments | | |
| Experiments | Sensitivity (90%CL) | Status |
| GEMMA | $\mu_\nu(\nu_e) <3\times10^{-11}\,\mu_B$ | 2004 |
| MAMONT | $\mu_\nu(\nu_e) <2\times10^{-12}\,\mu_B$ | R&D |
| Texono (ULEGe) | $\mu_\nu(\nu_e) <2\times10^{-11}\,\mu_B$ | R&D |

## Acknowledgments

The author would like to thank Dr. Jun Cao for his help. This work is supported in part by the Chinese Natural Science Foundation under the contract No. 10225524.

## References

1. Super-K Coll., Y. Fukuda *et al., Phys. Rev. Lett.* **81**, 1562(1998)
2. For a review, see C.K. Jung *et al*., *Ann. Rev. Nucl. Part. Sci.* **51**, 451(2001)
3. SNO Coll., Q.R. Ahmad, *Phys. Rev. Lett.* **87**, 071301(2001); *Phys. Rev. Lett.* **89**, 011301(2002)
4. KamLAND Coll., K. Eguchi *et al., Phys. Rev. Lett.* **90**, 021802(2003)
5. C. McGrew, this proceedings.
6. B.Pontecorvo, *Sov. Phys. JETP* **6**, 429 (1958); *Zh. Eksp. Theor. Fiz*.**53**,1717(1967); Z.Maki, M.Nakagawa and S. Sakata, *Prog.*

*Theor. Phys*. **28**,870(1962)
7. LSND Coll., C. Athanassopoulos *et al., Phys. Rev. Lett.* **81** 1774(1998)
8. Particle Data Group, E. Eidelman *et al., Phys. Lett. B* **592**, 1(2004).
9. Z.Z. Xing, Int. *J. Mod. Phys. A* **19**,1(2004).
10. M. Maltoni *et al.,* hep-ph/0405172
11. C. Saji, this proceedings; Super-K Coll., Y. Ashie *et al.,Phys.Rev.Lett.***93,**101801 (2004)
12. V.D.Barger *et al., Phys. Rev. Lett.* **82,** 2640 (1999)
13. E. Lisi *et al., Phys.Rev.Lett.* **85,** 1166(2000)
14. J. Bahcall and C. Pena-Garay404061*, New. J. Phys.* **6***,63(2004),* hep-ph/*0*404061.
15. H.H. Chen, *Phys. Rev. Lett.* **55**, 1534 (1985)
16. H. Deng, this proceedings; SNO Coll., A. Ahmed *et al., Phys. Rev. Lett .***92,** 181301 (2004)
17. A.Yu. Smirnov, talk given at "Neutrino 2002", hep-ph/0209131.
18. K. Ishihara, this proceedings; Super-K Coll., M.B. Smy *et al., Phys. Rev.D* **69**, 011104 (2004)
19. Super-K Coll., S. Fukuda *et al., Phys. Rev. Lett.* **86**, 5651 (2001); *Phys. Lett. B* **539**, 179(2002).
20. For a review, see C. Bemporad *et al., Rev. Mod. Phys*. **74***,97(2002)*
21. P. Vogel and J. Beacom, *Phys. Rev. D* **60***,* 053003 (1999)
22. I. Shimizu, this proceedings
23. KamLAND Coll., T. Araki *et al.,* hep-ex/0406035, submitted to *Phys. Rev. Lett.*
24. Chooz Coll., Apollinio et al., *Phys. Lett. B* **466**, 415(1999).
25. J.N. Bahcall *et al., JHEP* **0311***,004(2003),* hep-ph/0305159.
26. K. Anderson *et al.,* White paper report on using reactors to search for a value of $θ_{13}$, 2004,http://www.hep.anl.gov/minos/reactor13/white.html
27. L. Mikaelyan and V.V. Sinev, *Phys. Atom. Nucl.* **63**, 1002(2000), hep-ex/9908047.
28. Y.F. Wang *et al., Phys. Rev. D* **64**, 0013012 (2001); T. Hagner et al., *Astropart. Phys.* **14,** 33(2000)
29. K.B. Luk, this proceedings.
30. P. Huber et al., *Nucl. Phys. B***665**,487(2003)
31. Ch. Weinheimer, hep-ex/0306057.
32. A. Osipowicz *et al.,* hep-ex/0109033
33. H.V. Klapdor-Kleingrothaus *et al., Mod. Phys. Lett.* **16**,2409(2001); hep-ex/0205228; *Phys. Lett. B* **586**, 198(2004); C.E. Aalseth *et al., Mod. Phys. Lett.* **A17,** 1475(2002).
34. O. Elgaroy *et al.*, *Phys. Rev. Lett.* **89**, 061301 (2002)
35. D.N. Spergel *et al*., *Astrophys. J. Suppl.* **148**, 175(2003)
36. S. Hannestad, *JCPA* **0305**, 004 (2003).
37. S. Allen *et al., Mon. Not. Roy. Astron. Soc.* **346**, 593(2003).
38. M.Tegmark *et al.*, *Phys. Rev. D* **69**, 103501 (2004).
39. P.Crotty *et al.,Phys.Rev.D* **69**,123007(2004)
40. For a review, see S. Elliott and P. Vogel, *Annu. Rev. Nucl. Part. Sci.* **52**, 115(2002).
41. D. Lalanne, this proceedings.
42. R.Luescher et al.,*Phys.Lett.B* **434**,407(1998)
43. A.Alessandrello et al., *Phys. Lett. B* **486**, 13 (2000)
44. C.Arnaboldi et al., *Astropart. Phys.* **20**, 91 (2003)
45. S. Waldman, this proceedings.
46. H.V.Klapdor-Kleingrothaus, hep-ph/0103074.
47. R.Gaitskell et al., nucl-ex/0311013.
48. H.Ejiri et al., Phys. Rev. Lett.85,2917(2000)
49. Y. Takeuchi, this proceedings.
50. J. Li, this proceedings; H.B. Li *et al., Phys. Rev. Lett.* **90**, 131802 (2003)
51. MUNU Coll., Z. Daraktchieva *et al., Phys. Lett. B* **564**, 190(2003).
52. Super-K Coll., D.W. Liu *et al., Phys. Rev. Lett.* **93**, 021802 (2004)
53. Borexino Coll., H.O. Back *et al., Phys. Lett. B* **563**, 23(2003)
54. LSND Coll., L.B. Auerbach *et al., Phys. Rev. D* **63**, 112001(2001)
55. DONUT Coll., R. Schwienhorst *et al., Phys. Lett.* B **513**, 23(2001)